\documentclass[a4paper,11pt]{article}
\usepackage{jcappub} 
\usepackage{multirow}
\RequirePackage{graphicx}
\usepackage{amsmath}
\usepackage{graphicx}
\usepackage{subcaption}
\usepackage{tabularx}
\usepackage[a4paper, margin=2cm]{geometry}  

\title{\boldmath Testing Departures from the \(\Lambda\)CDM model in \(f(T)\) Gravity with DESI DR2}

\author[a,1]{Jie Zheng,\note{Corresponding author.}}
\author[b]{Mingzhi Wang}
\author[b]{and Ming Zhang}
\affiliation[a]{School of Microelectronics, Qingdao University of Science and Technology, Qingdao 266061, Shandong, China}

\affiliation[b]{School of Mathematics and Physics, Qingdao University of Science and Technology, Qingdao 266061, Shandong, China }

\emailAdd{zhengjie@mail.bnu.edu.cn}%

\abstract{Recent analyses combining DESI baryon acoustic oscillation (BAO) measurements with other cosmological observations have reported indications of departures from the $\Lambda$CDM model within dynamical dark energy scenarios. We investigate whether such deviations are reproduced consistently in $f(T)$ gravity and whether they persist when different cosmological observations are combined. Using the DESI DR2 BAO measurements, Cosmic Microwave Background (CMB), and Pantheon+ Type Ia Supernovae (SNe Ia) sample, we constrain three representative $f(T)$ models: the power-law, exponential, and hyperbolic-tangent models. We compare the constraints obtained from different combinations of data, distinguishing changes in marginalized parameter distributions from changes in the relative statistical performance of the models. For the power-law model, the mild deviation from the $\Lambda$CDM model obtained from BAO+CMB disappears after the SNe Ia data are included. In contrast, the exponential model retains a preference for a nonzero deformation parameter in the full analysis, while the allowed parameter space of the hyperbolic-tangent model changes substantially after the CMB distance priors are included. Importantly, the parameter-level deviation found for the exponential model does not lead to a clear statistical preference over the $\Lambda$CDM model according to the information criteria, whereas the hyperbolic-tangent model shows a poorer statistical performance for the combined data. Despite these differences in the model constraints, the inferred transition redshift $z_T$ remains relatively stable among the three models and data combinations. Our results show that the departures from the $\Lambda$CDM model indicated by DESI analyses are not a generic feature of $f(T)$ gravity, but depend strongly on the specific modified gravity parameterization and on the combination of cosmological observations.
\vspace{2mm}
}   

\begin{document}
\maketitle
\flushbottom

\section{Introduction}\label{sec:intro}

Over the past few decades, a variety of cosmological observations have established that the Universe is undergoing a phase of accelerated expansion~\cite{1998AJ....116.1009R,1999ApJ...517..565P}. In the standard $\Lambda$CDM model, this acceleration is driven by a cosmological constant, which provides the simplest description of dark energy (DE). Although $\Lambda$CDM has been remarkably successful in explaining a wide range of cosmological and astrophysical observations, it still faces both theoretical and observational challenges. On the theoretical side, the extremely small observed value of the cosmological constant compared with theoretical expectations gives rise to the well-known fine-tuning and cosmic coincidence problems~\cite{weinberg1,weinberg2}. On the observational side, discrepancies have emerged between cosmological parameters inferred from early- and late-Universe probes~\cite{Riess2022,Planck2018,2012MNRAS.427..146H,2020NatAs...4..196D,2021PhRvD.103d1301H,2021A&A...646A.140H,2021ApJ...908L...9D,2022PhRvD.105b3520A}. If these discrepancies cannot be fully attributed to systematic effects, they may point to physics beyond the standard $\Lambda$CDM framework and motivate the exploration of alternative descriptions of cosmic acceleration.
From a theoretical perspective, two main approaches have been developed to explain the accelerated expansion of the Universe. The first retains Einstein's general relativity and introduces additional dark energy components, while the second modifies the gravitational sector itself. Representative modified-gravity scenarios include the Dvali--Gabadadze--Porrati (DGP) braneworld model~\cite{Dvali:2000hr,Lian2021} and $f(R)$ gravity~\cite{Chiba:2003ir}. Of particular interest is $f(T)$ gravity~\cite{Bengochea:2008gz}, which extends the Teleparallel Equivalent of General Relativity (TEGR). In teleparallel gravity, gravitational interactions are described in terms of torsion rather than curvature: the torsion scalar $T$, constructed from the curvature-free Weitzenb\"ock connection, plays a role analogous to the Ricci scalar $R$ in general relativity. The $f(T)$ framework generalizes TEGR by replacing the torsion scalar in the gravitational action with a function $f(T)$, providing a geometrical mechanism for generating late-time cosmic acceleration.

The $f(T)$ framework has attracted considerable attention as a viable alternative description of cosmic acceleration for several reasons. First, it provides a geometrical mechanism for generating late-time acceleration without introducing an explicit cosmological constant, thereby offering a different perspective on the cosmological constant problem~\cite{Wu:2010mn,Nesseris:2013jea,Cai:2015emx,Cai:2019bdh}. Second, some viable $f(T)$ models can modify the late-time expansion history while remaining close to the standard cosmological evolution at high redshift, and have therefore been investigated in connection with the Hubble tension~\cite{Nunes:2016qyp,Bahamonde:2021gfp}. Moreover, the effective dark energy sector induced by modified teleparallel gravity can exhibit a time-dependent equation of state, including phantom-like behavior or crossing of the phantom divide~\cite{Bamba:2010wb,Cai:2015emx}. From a theoretical perspective, another appealing feature is that the field equations of $f(T)$ gravity remain second order, in contrast to the fourth-order equations generally encountered in metric $f(R)$ gravity, which simplifies both the theoretical and numerical treatment~\cite{Cai:2015emx,Krssak:2018ywd,Qi:2017xzl}. More detailed discussions of $f(T)$ gravity can be found in Refs.~\cite{Bengochea:2008gz,Cai:2015emx}.

The recent baryon acoustic oscillation (BAO) measurements from the Dark Energy Spectroscopic Instrument (DESI) provide a timely opportunity to revisit these modified-gravity scenarios. The DESI Data Releases 1 and 2 have substantially improved the precision of BAO measurements over a broad redshift range~\cite{2025JCAP...04..012A,DESI:2025zgx}. In particular, the DESI DR2 analysis, based on the first three years of observations and more than 14 million galaxies and quasars, provides significantly improved measurements of the cosmic distance--redshift relation. When combined with cosmic microwave background (CMB) and Type Ia supernovae (SNe Ia) observations, the DESI BAO measurements strengthen the preference for an evolving dark energy equation of state within parametrized dynamical dark energy models, although the significance depends on the adopted SNe Ia sample and model parameterization~\cite{Wang:2024pui,2024arXiv240704385L,Zheng:2024qzi,Lu:2024hvv,2025ApJ...987...58W,2026EPJC...86..924Z}. 

Although parametrized dynamical dark energy models, such as the Chevallier--Polarski--Linder (CPL) parameterization, provide a flexible description of departures from a cosmological constant, they remain phenomenological and do not by themselves specify the underlying physical mechanism responsible for cosmic acceleration. In this context, confronting viable $f(T)$ gravity models with the latest DESI DR2 measurements provides a useful test of whether the indications of departures from $\Lambda$CDM found in phenomenological dark energy analyses can also be accommodated within a modified-gravity framework. Several studies have begun to confront $f(T)$ gravity with the DESI DR2 measurements. In Ref.~\cite{Naik:2026DESI}, the power-law, Linder, and exponential $f(T)$ models were constrained using DESI DR2 BAO together with cosmic chronometers, DES Year 5 SNe Ia, and CMB distance priors. The analysis showed that deviations allowed by late-time observations are substantially reduced after the CMB information is included. More recently, Ref.~\cite{BouhmadiLopez:2026DESI} investigated several $f(T)$ models using Pantheon+, DESI DR2 BAO, CMB distance priors, and redshift-space distortion measurements, with particular emphasis on the $H_0$ and $S_8$ tensions and the growth of cosmic structure.
These studies demonstrate the constraining power of recent observations, while also motivating a closer comparison of how different distance probes affect the inferred model parameters. In particular, a displacement of a marginalized posterior away from the $\Lambda$CDM limit does not necessarily imply that the extended model is statistically preferred. The location and width of the posterior characterize parameter constraints within a model, whereas comparisons of the maximum likelihood and information criteria address its performance relative to a reference model. It is therefore useful to examine these complementary aspects together when assessing the response of $f(T)$ models to different combinations of cosmological observations.

In this work, we systematically investigate and compare three representative and widely studied $f(T)$ models: the power-law model $f(T)=\alpha(-T)^b$, the exponential model $f(T)=\alpha T_0[1-\exp(-p\sqrt{T/T_0})]$, and the hyperbolic tangent model $f(T)=\alpha(-T)^n\tanh(T_0/T)$, using the latest DESI DR2 BAO data, together with CMB distance priors and Pantheon+ SNe Ia. In particular, we assess their statistical performance relative to the $\Lambda$CDM model using the minimum $\chi^2$ and the information criteria, including the Akaike information criterion (AIC) and Bayesian information criterion (BIC), and further compare the transition redshift $z_T$ from cosmic deceleration to acceleration predicted by the different models. This combined statistical and dynamical analysis allows us to examine not only whether the additional degree of freedom in $f(T)$ gravity is supported by the latest cosmological observations, but also how different viable $f(T)$ scenarios modify the recent expansion history of the Universe. Our aim is to distinguish changes in parameter constraints from changes in model preference within a common set of distance observations.

The rest of the paper is organized as follows. In Sec.~\ref{sec:model}, we briefly introduce the three $f(T)$ models considered in this work. In Sec.~\ref{sec:data}, we describe the observational data and methodology used in our analysis. The main cosmological constraints, model-comparison results, and implications for the late-time expansion history are presented in Sec.~\ref{sec:results}. Finally, we summarize our conclusions in Sec.~\ref{sec:conclusion}.

\section{The $f(T)$ theory}
\label{sec:model}

We begin this section with a brief review of the cosmological formulation of $f(T)$ gravity and then introduce the three representative models considered in this work.

\subsection{The $f(T)$ cosmology}

In teleparallel gravity, the vierbein fields $e^{A}{}_{\mu}(x)$ ($A=0,1,2,3$) are used as the dynamical variables. They form an orthonormal basis for the tangent space at each point $x^\mu$ and satisfy
\begin{equation}
g_{\mu\nu}(x)
=
\eta_{AB}e^{A}{}_{\mu}(x)e^{B}{}_{\nu}(x),
\end{equation}
where $\eta_{AB}=\mathrm{diag}(1,-1,-1,-1)$ is the Minkowski metric.

Instead of the torsionless Levi--Civita connection used in general relativity, teleparallel gravity is formulated in terms of the curvature-free Weitzenb\"ock connection,
\begin{equation}
\hat{\Gamma}^{\lambda}{}_{\mu\nu}
=
e^{\lambda}{}_{A}\partial_{\nu}e^{A}{}_{\mu},
\end{equation}
which gives rise to a non-vanishing torsion tensor
\begin{equation}
T^{\lambda}{}_{\mu\nu}
=
e^{\lambda}{}_{A}
\left(
\partial_{\mu}e^{A}{}_{\nu}
-
\partial_{\nu}e^{A}{}_{\mu}
\right).
\end{equation}
The contortion tensor is defined as
\begin{equation}
K^{\rho}{}_{\mu\nu}
=
-\frac{1}{2}
\left(
T_{\mu}{}^{\rho}{}_{\nu}
-
T_{\nu}{}^{\rho}{}_{\mu}
-
T^{\rho}{}_{\mu\nu}
\right),
\end{equation}
and the superpotential is
\begin{equation}
S_{\rho}{}^{\mu\nu}
=
\frac{1}{2}
\left(
K^{\mu\nu}{}_{\rho}
+
\delta^{\mu}_{\rho}T^{\alpha\nu}{}_{\alpha}
-
\delta^{\nu}_{\rho}T^{\alpha\mu}{}_{\alpha}
\right).
\end{equation}
The torsion scalar is then constructed as
\begin{equation}
T
=
S_{\rho}{}^{\mu\nu}T^{\rho}{}_{\mu\nu}.
\label{eq:torsion_scalar}
\end{equation}

In the teleparallel equivalent of general relativity (TEGR), the gravitational Lagrangian is given by the torsion scalar $T$. Throughout this work, $f(T)$ denotes the modification to the TEGR Lagrangian, while the complete gravitational Lagrangian is denoted by $\mathcal{F}(T)=T+f(T)$.
The corresponding action in the presence of matter is therefore
\begin{equation}
S
=
\frac{1}{16\pi G}
\int d^{4}x\,e\,[T+f(T)]
+
\int d^{4}x\,e\,\mathcal{L}_{m}.
\end{equation}
where $e\equiv\det(e^{A}{}_{\mu})=\sqrt{-g}$ and $\mathcal{L}_{m}$ denotes the matter Lagrangian density. The TEGR limit is recovered for $f(T)=0$.

Varying the action with respect to the vierbein gives
\begin{align}
&
\left[
e^{-1}\partial_{\mu}
\left(e S_{A}{}^{\mu\nu}\right)
-
e_{A}{}^{\lambda}
T^{\rho}{}_{\mu\lambda}
S_{\rho}{}^{\nu\mu}
\right]
\left(1+f_{T}\right)
\nonumber\\
&\quad
+
S_{A}{}^{\mu\nu}
(\partial_{\mu}T)f_{TT}
+
\frac{1}{4}
e_{A}{}^{\nu}
\left[T+f(T)\right]
=
4\pi G\,
e_{A}{}^{\rho}
\mathcal{T}_{\rho}{}^{\nu},
\label{eq:field_equation_fT}
\end{align}
where $f_{T}\equiv df/dT$, $f_{TT}\equiv d^{2}f/dT^{2}$, and $\mathcal{T}_{\rho}{}^{\nu}$ is the matter energy-momentum tensor.

For a spatially flat Friedmann-Robertson-Walker (FRW) Universe, we adopt the diagonal vierbein
\begin{equation}
e^{A}{}_{\mu}
=
\mathrm{diag}(1,a,a,a),
\end{equation}
which gives the spatially flat FRW metric
\begin{equation}
ds^{2}
=
dt^{2}
-
a^{2}(t)\delta_{ij}dx^{i}dx^{j}.
\end{equation}
The torsion scalar then reduces to
\begin{equation}
T=-6H^{2},
\label{eq:T_FRW}
\end{equation}
where $H\equiv\dot{a}/a$ is the Hubble parameter.

Assuming that the matter sector is described by a perfect fluid with total energy density $\rho$ and pressure $p$, the modified Friedmann equations can be written as
\begin{align}
H^{2}
&=
\frac{8\pi G}{3}\rho
-\frac{f}{6}
+\frac{T f_{T}}{3},
\label{eq:Friedmann1_fT}
\\
\dot{H}
&=
-\frac{4\pi G(\rho+p)}
{1+f_{T}+2T f_{TT}}.
\label{eq:Friedmann2_fT}
\end{align}

The modification to TEGR can be interpreted as an effective dark energy component. Its effective energy density and pressure are given by
\begin{align}
\rho_{\rm eff}
&=
\frac{1}{16\pi G}
\left(
2T f_{T}-f
\right),
\label{eq:rho_eff}
\\
p_{\rm eff}
&=
\frac{1}{16\pi G}
\left[
4\dot{H}
\left(
f_{T}+2T f_{TT}
\right)
-2T f_{T}
+f
\right],
\label{eq:p_eff}
\end{align}
and the corresponding effective equation-of-state parameter is
\begin{equation}
w_{\rm eff}
=
\frac{p_{\rm eff}}{\rho_{\rm eff}}.
\label{eq:w_eff}
\end{equation}

Introducing the normalized Hubble parameter
\begin{equation}
E(z)\equiv\frac{H(z)}{H_{0}},
\end{equation}
and using
\begin{equation}
T=T_{0}E^{2}(z),
\qquad
T_{0}=-6H_{0}^{2},
\end{equation}
the background expansion equation can be expressed as
\begin{equation}
E^{2}(z,\mathbf{p})
=
\Omega_{m}(1+z)^{3}
+
\Omega_{r}(1+z)^{4}
+
\Omega_{F}\,y(z,\mathbf{p}),
\label{eq:E2_general}
\end{equation}
where
\begin{equation}
\Omega_{F}
=
1-\Omega_{m}-\Omega_{r},
\end{equation}
and
\begin{equation}
\Omega_{i}
=
\frac{8\pi G\rho_{i0}}
{3H_{0}^{2}},
\qquad i=m,r.
\end{equation}
The function $y(z,\mathbf{p})$ characterizes the modification to the standard expansion history and is defined as
\begin{equation}
y(z,\mathbf{p})
=
\frac{
f(T)-2T f_{T}(T)
}{
T_{0}\Omega_{F}
},
\label{eq:y_general}
\end{equation}
where $\mathbf{p}$ denotes the free parameter or parameters of a specific $f(T)$ model. The normalization condition $E(z=0)=1$ implies
\begin{equation}
y(0,\mathbf{p})=1.
\end{equation}

\subsection{Representative \texorpdfstring{$f(T)$}{f(T)} models}
\label{subsec:fT_models}

In this work, we consider three specific $f(T)$ models that have been widely studied and have passed basic observational tests. For each model, the normalization parameter is determined by imposing the boundary condition $E(z=0)=1$. According to the above convention, the functions $f_i(T)$ introduced below denote the modifications to TEGR, while the corresponding total gravitational Lagrangian is $\mathcal{F}_i(T)=T+f_i(T)$.

\subsubsection{The power-law model ($f_1$CDM)}

The first model is the power-law model \cite{Bengochea:2008gz},
\begin{equation}
f_1(T)
=
\alpha(-T)^{b},
\label{eq:fT_powerlaw}
\end{equation}
where $b$ quantifies the deviation from the $\Lambda$CDM model, while
$\alpha$ is fixed by the present-day normalization condition. Substituting
Eq.~(\ref{eq:fT_powerlaw}) into the background equation gives
\begin{equation}
\alpha
=
(6H_{0}^{2})^{1-b}
\frac{\Omega_{F}}
{2b-1}.
\label{eq:alpha_powerlaw}
\end{equation}
The corresponding deformation function is
\begin{equation}
y_1(z;b)
=
E^{2b}(z).
\label{eq:y_powerlaw}
\end{equation}
The normalized Hubble parameter therefore satisfies
\begin{equation}
E^{2}(z)
=
\Omega_{m}(1+z)^{3}
+
\Omega_{r}(1+z)^{4}
+
\Omega_{F}E^{2b}(z).
\label{eq:E2_powerlaw}
\end{equation}
The $\Lambda$CDM model is recovered for $b=0$.

\subsubsection{The exponential model ($f_2$CDM)}

The second model is the exponential model \cite{Linder:2010py},
\begin{equation}
f_2(T)
=
\alpha T_{0}
\left[
1-
\exp\left(
-p\sqrt{\frac{T}{T_{0}}}
\right)
\right],
\label{eq:fT_exponential}
\end{equation}
where $\alpha$ and $p$ are dimensionless parameters. Imposing
$E(z=0)=1$ gives
\begin{equation}
\alpha
=
\frac{\Omega_{F}}
{
1-(1+p)e^{-p}
}.
\label{eq:alpha_exponential}
\end{equation}
The corresponding deformation function is
\begin{equation}
y_2(z;p)
=
\frac{
1-\left[1+pE(z)\right]e^{-pE(z)}
}{
1-(1+p)e^{-p}
}.
\label{eq:y_exponential_p}
\end{equation}
This model approaches the $\Lambda$CDM limit as
$p\rightarrow+\infty$.

Following the parameterization commonly adopted for this model, we
introduce
\begin{equation}
b\equiv\frac{1}{p}.
\end{equation}
The deformation function can then be rewritten as
\begin{equation}
y_2(z;b)
=
\frac{
1-
\left(
1+\dfrac{E(z)}{b}
\right)
\exp\left[-\dfrac{E(z)}{b}\right]
}{
1-
\left(
1+\dfrac{1}{b}
\right)
\exp\left(-\dfrac{1}{b}\right)
}.
\label{eq:y_exponential_b}
\end{equation}
In terms of $b$, the $\Lambda$CDM limit corresponds to
$b\rightarrow0^{+}$.

The corresponding normalized Hubble parameter is
\begin{equation}
E^{2}(z)
=
\Omega_{m}(1+z)^{3}
+
\Omega_{r}(1+z)^{4}
+
\Omega_{F}\,y_2(z;b).
\label{eq:E2_exponential}
\end{equation}

\subsubsection{The hyperbolic-tangent model ($f_3$CDM)}

The third model is the hyperbolic-tangent model \cite{Wu:2010av},
\begin{equation}
f_3(T)
=
\alpha(-T)^{n}
\tanh\left(\frac{T_{0}}{T}\right),
\label{eq:fT_tanh}
\end{equation}
where $\alpha$ and $n$ are the model parameters. The normalization
condition $E(z=0)=1$ gives
\begin{equation}
\alpha
=
-\frac{
\Omega_{F}(6H_{0}^{2})^{1-n}
}{
2\,\mathrm{sech}^{2}(1)
+
(1-2n)\tanh(1)
}.
\label{eq:alpha_tanh}
\end{equation}
The corresponding deformation function is
\begin{equation}
y_3(z;n)
=
E^{2(n-1)}(z)
\frac{
2\,\mathrm{sech}^{2}\!\left[E^{-2}(z)\right]
+
(1-2n)E^{2}(z)
\tanh\!\left[E^{-2}(z)\right]
}{
2\,\mathrm{sech}^{2}(1)
+
(1-2n)\tanh(1)
}.
\label{eq:y_tanh}
\end{equation}
Hence, the normalized Hubble parameter satisfies
\begin{equation}
E^{2}(z)
=
\Omega_{m}(1+z)^{3}
+
\Omega_{r}(1+z)^{4}
+
\Omega_{F}\,y_3(z;n).
\label{eq:E2_tanh}
\end{equation}

Unlike the $f_1$CDM and $f_2$CDM models, the $f_3$CDM model cannot be
reduced to the $\Lambda$CDM model for any value of $n$. Consequently,
$n$ does not directly quantify the deviation from $\Lambda$CDM.
Furthermore, requiring a positive effective energy density leads to
\begin{equation}
n>\frac{3}{2}.
\label{eq:n_tanh_condition}
\end{equation}

The three models therefore represent qualitatively different
modifications of the teleparallel gravitational sector. The $f_1$CDM
and $f_2$CDM models contain $\Lambda$CDM as a limiting case, whereas
$f_3$CDM represents a genuinely distinct background expansion history.

\section{Data and Methodology}\label{sec:data}
In this section, we describe the observational datasets and the analysis methodology adopted in this work. We briefly introduce the three cosmological probes employed in our analysis and then outline the statistical framework used to constrain the $f(T)$ models with different combinations of BAO, BAO+CMB, and BAO+CMB+SN data.

\subsection{DESI BAO DR2 Measurement}

The Dark Energy Spectroscopic Instrument (DESI) is conducting a Stage IV spectroscopic survey designed to substantially improve cosmological constraints. The survey spans approximately $14{,}200$ square degrees over a five-year observational campaign and covers a broad redshift range of $0.1<z<4.2$, with a spectroscopic sample size roughly an order of magnitude larger than that of previous surveys.
Our analysis adopts the latest DESI DR2 BAO measurements reported in Ref.~\cite{DESI:2025zgx}. The compressed BAO likelihood adopted in our analysis contains 13 observables: the isotropic measurement $D_V/r_d$ at $z_{\rm eff}=0.295$ from the BGS sample, together with the transverse and radial measurements, $D_M/r_d$ and $D_H/r_d$, at $z_{\rm eff}=0.510$, $0.706$, $0.934$, $1.321$, $1.484$, and $2.330$. The full covariance between the transverse and radial measurements is included in the likelihood analysis.

The DESI DR2 measurements are expressed in terms of three key cosmological distance ratios. The first observable is
\begin{equation}
\label{eq:DM}
\frac{D_{M}(z)}{r_{d}}=\frac{D_{L}(z)}{r_{d}(1+z)},
\end{equation}
where $D_{M}(z)$ denotes the transverse comoving distance and $r_d$ is the sound horizon at the baryon drag epoch. The value of $r_d$ depends on the physical matter and baryon densities, as well as the effective number of relativistic species. In our analysis, the combination $r_d h$ is treated as a free nuisance parameter.
For a given cosmological model characterized by parameters $\hat{p}$, the luminosity distance is given by
\begin{equation}
D_{\mathrm{L}}(z,\hat{p})=\frac{c(1+z)}{H_{0}}\int^{z}_{0} \frac{dz^{\prime}}{E(z^{\prime})},
\end{equation}
where $E(z)=H(z)/H_0$ is the normalized Hubble parameter.
The second observable is
\begin{equation}
\label{eq:DH}
\frac{D_{H}(z)}{r_{d}}=\frac{c}{r_{d}H(z)}.
\end{equation}
The third observable is the volume-averaged distance,
\begin{equation}
\label{eq:DV}
\frac{D_V(z)}{r_d}
=
\frac{
\left[
zD_M^2(z)D_H(z)
\right]^{1/3}}{r_d}.
\end{equation}

Then, the $\chi^{2}$ likelihood function of BAO data is expressed as
\begin{equation}
\chi^{2}_{\mathrm{BAO}}=\Delta \mathrm{X}^{T}  \mathbf{C}_{\mathrm{BAO}}^{-1} \Delta \mathrm{X},
\end{equation}
where $\Delta\mathrm{X} =\mathrm{X}_{\rm obs}-\mathrm{X}_{\rm th}$ denotes the residual vector between the DESI DR2 measurements and the corresponding theoretical predictions. The covariance matrix $\mathbf{C}_{\rm BAO}$ is taken from the DESI DR2 data release \cite{DESI:2025zgx}.

\subsection{Cosmic Microwave Background (CMB)}
Precise measurements of the CMB temperature and polarization anisotropies provide crucial insights into the physics of the early Universe and the formation of large-scale structure. As a fundamental cosmological probe, CMB observations encode a wealth of information on key cosmological parameters \cite{2011ApJS..192...18K,2013ApJS..208...19H,2014A&A...571A..16P,Planck2018}, including the Hubble constant and the properties of dark energy.

Following Refs.~\cite{wang2007prd_dp,2019JCAP...02..028C}, we adopt the Planck 2018 CMB distance priors rather than the full CMB power-spectrum likelihood. This compressed approach retains the dominant geometrical information relevant to the background expansion while avoiding the computational complexity of a full CMB power-spectrum likelihood analysis.

The CMB distance prior information adopted in our analysis consists of the shift parameter $R$, the acoustic scale $l_A$, and the physical baryon density $\Omega_bh^2$. These quantities provide robust constraints on the background expansion history and play an important role in our joint cosmological analysis.
The acoustic scale and the shift parameter are defined as
\begin{equation}
    \label{eq:la}
    l_{A}=(1+z_{*})\frac{\pi D_{A}(z_{*})}{r_{s}(z_{*})},
\end{equation}
\begin{equation}
\label{eq:R}
R
\equiv
\frac{H_0\sqrt{\Omega_m}}{c}
(1+z_*)D_A(z_*).
\end{equation}
where $z_{*}$ is the redshift at the photon decoupling epoch, which can be approximately calculated by
\begin{equation}
    \label{eq:z*}
    z_{*}=1048[1+0.00124(\Omega_{b}h^{2})^{-0.738}][1+g_{1}(\Omega_{m}h^{2})^{g_{2}}],
\end{equation}
where 
\begin{equation}
    g_{1}=\frac{0.0783(\Omega_{b}h^{2})^{-0.238}}{1+39.5(\Omega_{b}h^{2})^{0.763}},
\end{equation}
\begin{equation}
    g_{2}=\frac{0.560}{1+21.1(\Omega_{b}h^{2})^{1.81}}.
\end{equation}
In Eq.~(\ref{eq:la}), the comoving sound horizon is defined by
\begin{equation}
    \label{eq:rs}
    r_{s}(z)=\frac{c}{H_{0}}\int^{a}_{0} \frac{da}{a^{2}E(a)\sqrt{3(1+\frac{3\Omega_{b}h^{2}}{4\Omega_{\gamma}h^{2}}a)}},
\end{equation}
where $a=1/(1+z)$, $\frac{3}{4\Omega_{\gamma}h^{2}}=31500(T_{CMB}/2.7K)^{-4}$ and $T_{CMB}=2.7255K$.
And the angular diameter distance $D_{A}$ is written by
\begin{equation}
\label{eq:DA}
D_A(z)
=
\frac{c}{H_0(1+z)}
\int_0^z
\frac{dz^{\prime}}{E(z^{\prime})}.
\end{equation}

The $\chi^{2}$ of the CMB is given by
\begin{equation}
    \chi^{2}_{CMB}=\Delta \mathrm{Y}^{T} \mathbf{C}_{\mathrm{CMB}}^{-1} \Delta \mathrm{Y},
\end{equation}
where $\Delta \mathrm{Y}=y_{i}-y_{\mathrm{model}}$, with \begin{equation}
\mathbf{y_i}
=
\begin{pmatrix}
1.7502 \\
301.471 \\
0.02236
\end{pmatrix},
\end{equation}
\begin{equation}
\mathbf{y}_{\rm model}
=
\begin{pmatrix}
R \\
l_A \\
\Omega_b h^2
\end{pmatrix},
\end{equation}\
The corresponding inverse covariance matrix $\mathbf{C}^{-1}_{\mathrm{CMB}}$ is taken from Ref.~\cite{2019JCAP...02..028C}.

\subsection{Type Ia Supernovae (SNe Ia)}

Type Ia supernovae are luminous and well-calibrated standardizable candles that play a pivotal role in modern cosmology, particularly in probing the late-Universe expansion history. Their well-characterized light-curve properties and relatively uniform peak luminosities enable precise distance measurements, making them powerful probes for constraining cosmological parameters. In this work, we adopt the full Pantheon+ compilation \cite{pantheonp}, which consists of 1,701 high-quality light curves from 1,550 spectroscopically confirmed SNe Ia spanning a wide redshift range of $0.00122 < z < 2.26137$. To minimize potential systematic effects associated with very low-redshift observations, we exclude data points with $z < 0.01$. After applying this cut, the final sample contains 1590 data points.

For this compilation, the observed distance modulus is given by
\begin{equation}
\mu_{\mathrm{obs,i}}=m_{B}+\alpha x_{1}-\beta c-M-\delta_{\mathrm{bias}}+\delta_{\mathrm{host}},
\end{equation}
where $m_{B}$ denotes the apparent B-band magnitude, $x_{1}$ is the stretch parameter corresponding to light-curve width, $c$ is the light-curve color that includes contributions from both intrinsic color and dust, $M$ is the absolute B-band magnitude of a fiducial SN Ia and treated as a nuisance parameter in our analysis, $\delta_{\mathrm{bias}}$ is a correction term to account for selection biases, and $\delta_{\mathrm{host}}$ is the luminosity correction for residual correlations. $\alpha$ and $\beta$ are global nuisance parameters that relate stretch and color to luminosity, respectively. 
The theoretical distance modulus $\mu_{\mathrm{th}}$ is defined as 
\begin{equation}
    \mu_{\mathrm{th}}(z)= 5\mathrm{log}_{10}[\frac{D_{L}(z)}{\mathrm{Mpc}}]+25,
\end{equation}
where $D_{\mathrm{L}}(z)$ is the luminosity distance, 
\begin{equation}
\label{eq:DL}
   D_L(z)=\frac{c(1+z)}{H_0}\int^{z}_{0} \frac{d \tilde{z}}{E(\tilde{z})}.
\end{equation}

Then, the $\chi^{2}$ likelihood function of Pantheon+ compilation can be written as
\begin{equation}
\chi_{\mathrm{SN}}^{2}=\Delta \mathrm{D}^{T} \mathbf{C}_{\mathrm{SN}}^{-1} \Delta \mathrm{D},
\end{equation}
where the covariance matrix $\mathbf{C}$ is the covariance matrix including both the systematic and statistical errors, which can be found on the website\footnote{https://github.com/PantheonPlusSH0ES/DataRelease}, and $\Delta \vec{D}$ is the vector of SNe Ia distance modulus residuals,
\begin{equation}
    \Delta \mathrm{D}_{i}=\mu_{\mathrm{obs,i}}-\mu_{\mathrm{model}}(z_{i}).
\end{equation}

\section{Results and Discussion}\label{sec:results}

In this section, we present the observational constraints on the three $f(T)$ cosmological models considered in this work. Recent analyses combining DESI BAO with other cosmological observations have reported indications of departures from the $\Lambda$CDM model within dynamical dark energy scenarios. Motivated by these results, we examine whether similar deviations are also allowed by the DESI DR2 BAO measurements in the framework of $f(T)$ gravity. Then, we examine how the corresponding parameter regions change when the CMB distance priors and the Pantheon+ SNe Ia sample are progressively included. It also allows us to examine whether the indication of a deviation from the $\Lambda$CDM model remains in the combined analysis and whether the result depends on the specific $f(T)$ model. We further investigate the transition redshift $z_T$ to characterize the onset of late-time cosmic acceleration and assess the relative statistical performance of the models using the Akaike information criterion (AIC) and the Bayesian information criterion (BIC).

\begin{table*}[htbp]
\centering
\renewcommand{\arraystretch}{2}                                          

\begin{tabular}{|l|c|c|c|c|}
\hline
\textbf{Model} & \textbf{Data} & $\Omega_{m}$ & $b$ or $n$  & $z_T$ \\ \hline
\multirow{3}{*}{$f_1\mathrm{CDM}\  \mathrm{model}$} 
& BAO   &  $0.29\pm0.01$ &  $0.14^{+0.15}_{-0.16}$  & $0.67^{+0.05}_{-0.04}$ \\ \cline{2-5}
& BAO+CMB & $0.30\pm0.01$  & $-0.08\pm0.07$ & $0.62^{+0.04}_{-0.03}$ \\ \cline{2-5}
& BAO+CMB+SN & $0.31\pm0.01$  & $0.00\pm0.05$  & $0.65\pm0.03$ \\ \hline
\multirow{3}{*}{$f_2\mathrm{CDM} \  \mathrm{model}$} 
& BAO  & $0.31\pm0.01$  &  $0.27^{+0.12}_{-0.17}$  & $0.68\pm0.03$ \\ \cline{2-5}
& BAO+CMB&  $0.31\pm0.01$ & $0.13^{+0.10}_{-0.09}$  &  $0.65\pm0.03$\\ \cline{2-5}
& BAO+CMB+SN  &  $0.29\pm0.01$ & $0.35^{+0.08}_{-0.07}$  & $0.70\pm0.03$\\ \hline
\multirow{3}{*}{$f_3\mathrm{CDM} \  \mathrm{model}$} 
& BAO &  $0.22^{+0.03}_{-0.04}$ & $1.86^{+0.07}_{-0.08}$  & $0.70^{+0.14}_{-0.16}$ \\ \cline{2-5}
& BAO+CMB & $0.30\pm0.01$  & $1.60\pm0.01$  & $0.67^{+0.02}_{-0.03}$\\ \cline{2-5}
& BAO+CMB+SN &  $0.31\pm0.01$ &  $1.62\pm0.01$  & $0.68^{+0.02}_{-0.03}$ \\ \hline
\end{tabular}
\caption{
Marginalized constraints on the cosmological parameters and the transition redshift for the three $f(T)$ models considered in this work, obtained from BAO, BAO+CMB, and BAO+CMB+SN data. The quoted uncertainties correspond to the 68\% credible interval. For the $f_1$CDM and $f_2$CDM models, $b$ denotes the model parameter, while $n$ is the corresponding parameter of the $f_3$CDM model. The transition redshift $z_T$ is defined by $q(z_T)=0$.}
\label{tab:results}
\end{table*}

\begin{table*}[htbp]
\centering
\renewcommand{\arraystretch}{2}                                          

\begin{tabular}{|l|c|c|c|c|c|c|}
\hline
\textbf{Model} & \textbf{Data} & $\chi_{\mathrm{min}}^2
$ & AIC & $\Delta \mathrm{AIC}$ & BIC & $\Delta \mathrm{BIC}$   \\ \hline
\multirow{3}{*}{$\Lambda \mathrm{CDM}\  \mathrm{model}$} 
& BAO   &  10.52 &  14.52  & 0  &  15.65  & 0  \\ \cline{2-7}
& BAO+CMB & 13.34  & 21.34  & 0 &  24.43  & 0 \\ \cline{2-7}
& BAO+CMB+SN & 1417.99 & 1427.99  & 0 & 1454.89 & 0 \\ \hline
\multirow{3}{*}{$f_1\mathrm{CDM}\  \mathrm{model}$} 
& BAO   &  9.86 &  15.86  &  +1.34 &  17.55 & +1.90  \\ \cline{2-7}
& BAO+CMB & 11.78  & 21.78  &  +0.44 & 25.64 & +1.21\\ \cline{2-7}
& BAO+CMB+SN & 1417.99  & 1429.99 & +2.00 &  1462.28 & +7.39 \\ \hline
\multirow{3}{*}{$f_2\mathrm{CDM}\ \mathrm{model}$} 
& BAO  & 9.07  &  15.07 & +0.55 & 16.76 & +1.11 \\ \cline{2-7}
& BAO+CMB&  13.09  & 23.09  & +1.75 & 26.96 & +2.53 \\ \cline{2-7}
& BAO+CMB+SN  &  1415.89 & 1427.89  & -0.10  & 1460.18 & +5.29 \\ \hline
\multirow{3}{*}{$f_3\mathrm{CDM}\ \mathrm{model}$} 
& BAO &  10.85 & 16.85  & +2.33 & 18.54 & +2.89 \\ \cline{2-7}
& BAO+CMB & 22.49  & 32.49  & +11.15 & 36.35  & +11.92 \\ \cline{2-7}
& BAO+CMB+SN & 1435.16 &  1447.16  &  +19.17 & 1479.45 & +24.56 \\ \hline
\end{tabular}
\caption{
Minimum chi-square values and the corresponding Akaike information
criterion (AIC) and Bayesian information criterion (BIC) for the
$\Lambda$CDM, $f_1$CDM, $f_2$CDM, and $f_3$CDM models using different
data combinations. The quantities $\Delta{\rm AIC}$ and $\Delta{\rm BIC}$
are evaluated relative to the $\Lambda$CDM model for the same data
combination, such that $\Delta{\rm AIC}=\Delta{\rm BIC}=0$ for the
$\Lambda$CDM model. Smaller values of AIC and BIC indicate a more favorable
balance between goodness of fit and model complexity.
}
\label{tab:IC}
\end{table*}

\begin{figure}[htbp]
    \centering
    \includegraphics[width=0.7\linewidth]{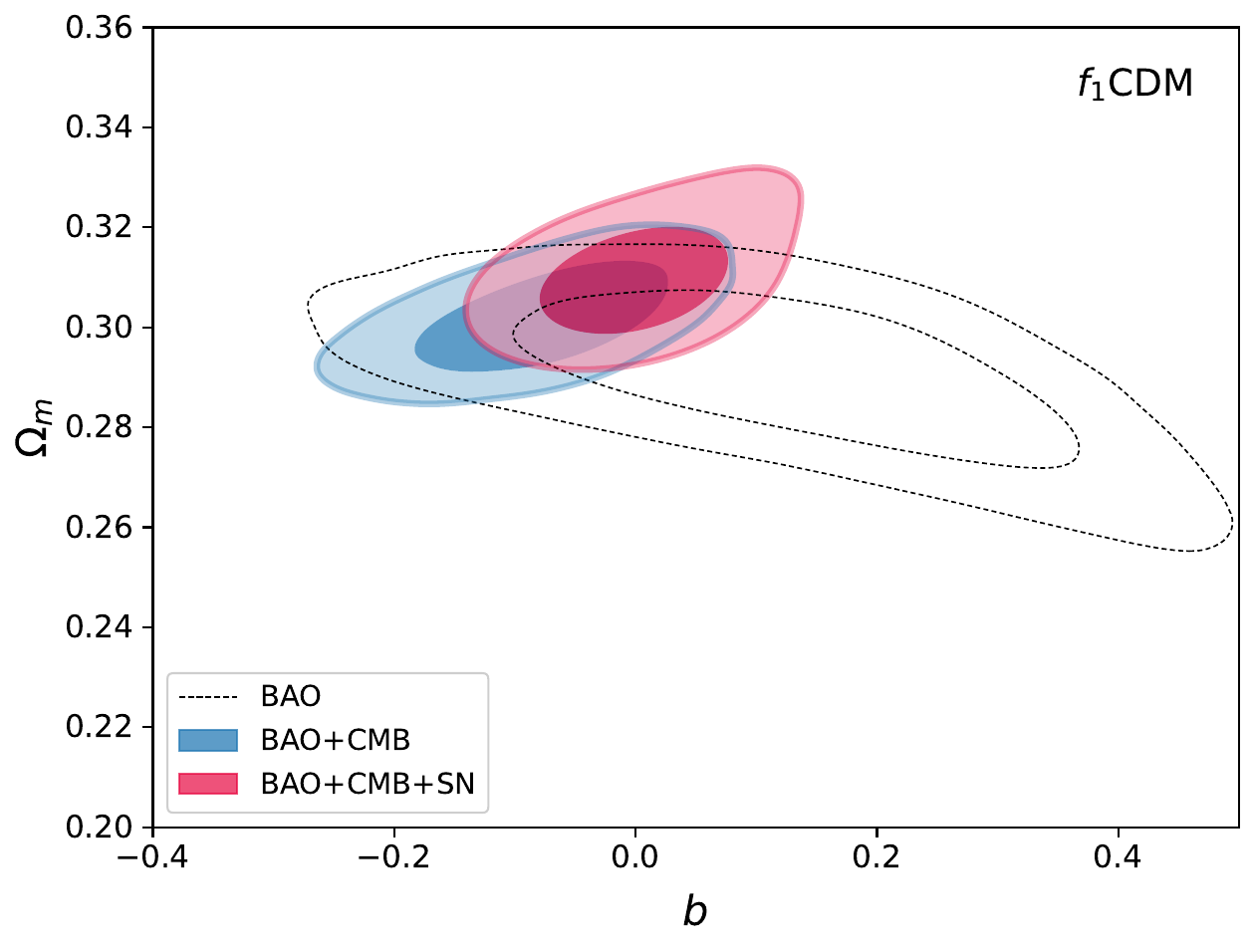}
    \caption{
Two-dimensional marginalized constraints in the $b$--$\Omega_m$ plane
for the $f_1$CDM model obtained from BAO, BAO+CMB, and BAO+CMB+SN data.
The inner and outer contours correspond to the 68\% and 95\% credible regions, respectively.
    }
    \label{fig:f1}
\end{figure}

\begin{figure}[htbp]
    \centering
    \includegraphics[width=0.7\linewidth]{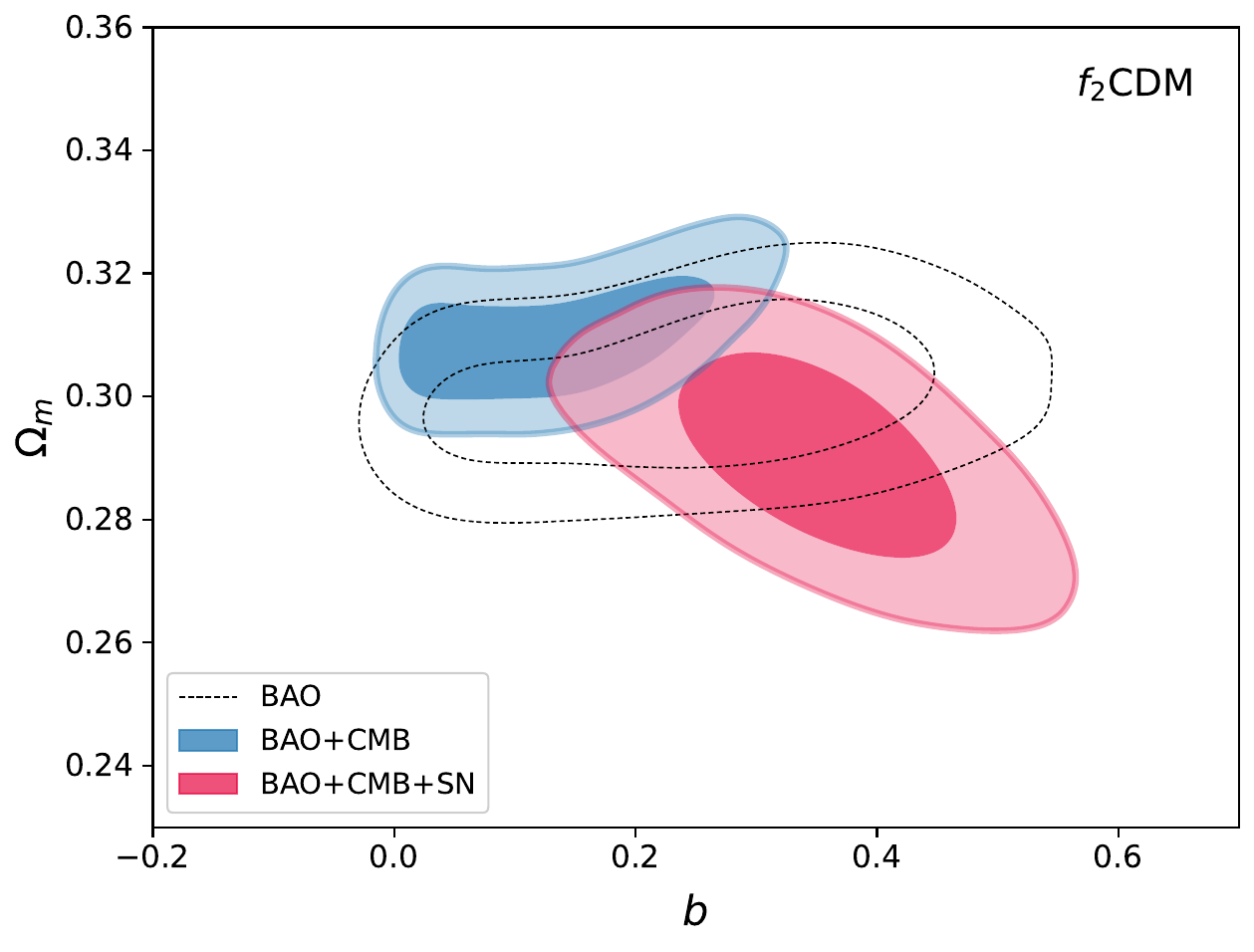}
    \caption{
Two-dimensional marginalized constraints in the $b$--$\Omega_m$ plane
for the $f_2$CDM model obtained from BAO, BAO+CMB, and BAO+CMB+SN data.
The inner and outer contours correspond to the 68\% and 95\% credible regions, respectively.
    }
    \label{fig:f2}
\end{figure}

\begin{figure}[htbp]
    \centering
    \includegraphics[width=0.7\linewidth]{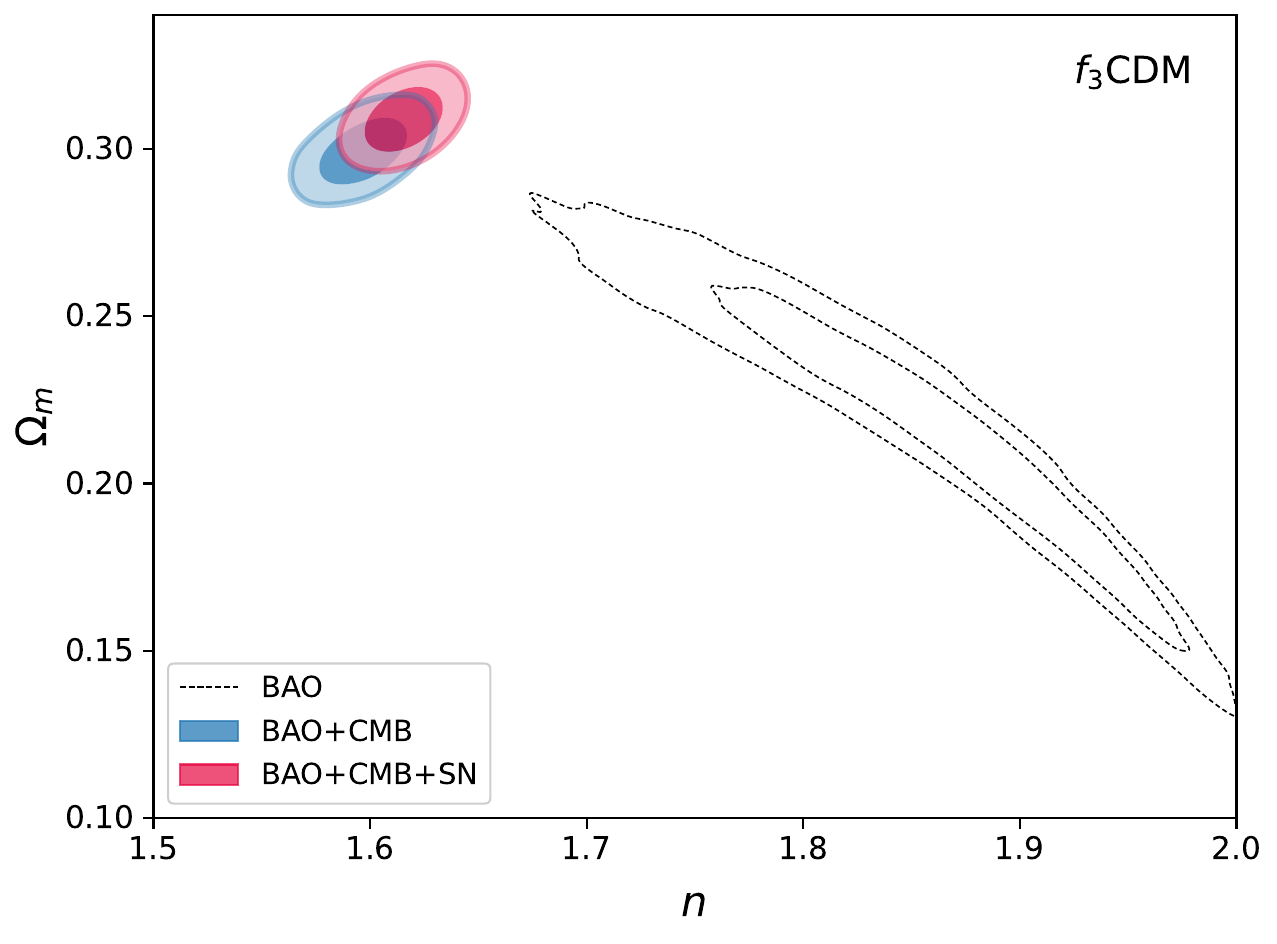}
    \caption{
Two-dimensional marginalized constraints in the $n$--$\Omega_m$ plane
for the $f_3$CDM model obtained from BAO, BAO+CMB, and BAO+CMB+SN data.
The inner and outer contours correspond to the 68\% and 95\% credible regions, respectively.
    }
    \label{fig:f3}
\end{figure}

\begin{figure}[htbp]
    \centering
    \includegraphics[width=0.7\linewidth]{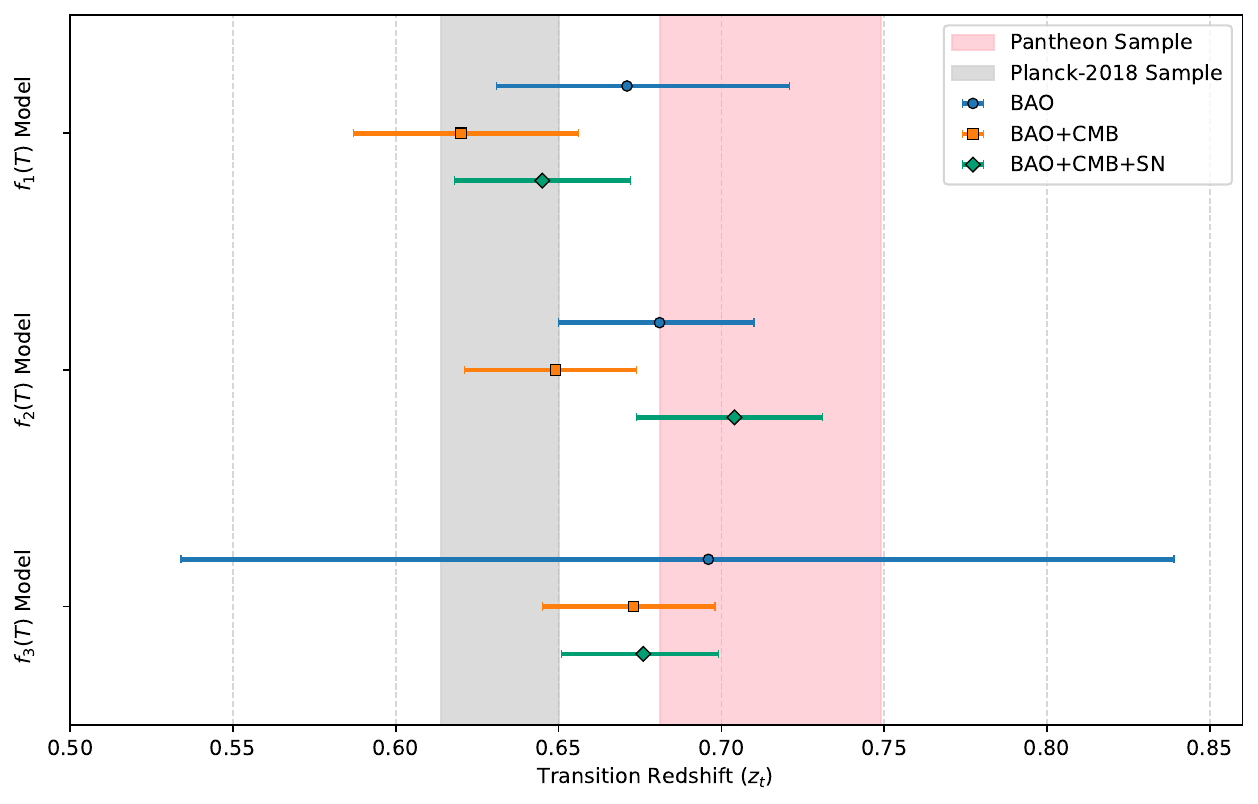}
    \caption{Constraints on the transition redshift $z_T$ for the three $f(T)$ models obtained from the BAO, BAO+CMB, and BAO+CMB+SN data combinations. The points and error bars denote the marginalized constraints at the 68\% credible interval. The gray and pink shaded regions represent the reference ranges obtained from the Planck 2018 and Pantheon samples, respectively, following Ref.~\cite{Kumar:2025mzo}.}    
    \label{fig:zt}
\end{figure}

For the $f_1$CDM model, the BAO-only analysis gives $b=0.14^{+0.15}_{-0.16}$ and $\Omega_m=0.29\pm0.01$, and the result is consistent with the $\Lambda$CDM model. After including the CMB distance priors, the constraint becomes $b=-0.08\pm0.07$, while $\Omega_m$ remains close to $0.30$. The uncertainty on $b$ is substantially reduced, and $b=0$ lies slightly outside the 68\% credible interval. However, this behavior changes when the Pantheon+ SNe Ia sample is added. For the full combination, we obtain $b=0.00\pm0.05$ and $\Omega_m=0.31\pm0.01$. The preferred value of $b$ is therefore fully consistent with the $\Lambda$CDM model. The mild deviation found in the BAO+CMB case does not persist once the SNe Ia data are included. This behavior is particularly interesting in the context of recent DESI analyses of dynamical dark energy, where combinations of DESI BAO, CMB, and SNe Ia can retain a preference for departures from the $\Lambda$CDM model. In the $f_1$CDM model considered here, by contrast, the full data combination is consistent with the $\Lambda$CDM model. This result indicates that the departure from the $\Lambda$CDM model suggested in dynamical dark energy analyses is not reproduced in the same way in the $f_1(T)$ model.

A different behavior is found for the $f_2$CDM model. The BAO-only analysis gives $b=0.27^{+0.12}_{-0.17}$ and $\Omega_m=0.31\pm0.01$. Since $b=0$ corresponds to the $\Lambda$CDM model, the BAO-only result shows a mild preference for a positive value of $b$, with $b=0$ lying outside the 68\% credible interval. After including the CMB distance priors, the constraint becomes $b=0.13^{+0.10}_{-0.09}$, while $\Omega_m$ remains nearly unchanged. The preferred value of $b$ therefore moves closer to zero, indicating that the inclusion of the CMB information reduces the deviation from the $\Lambda$CDM model.
The trend changes again when the Pantheon+ SNe Ia sample is added. For the full BAO+CMB+SN combination, we obtain $b=0.35^{+0.08}_{-0.07}$ and $\Omega_m=0.29\pm0.01$. The preferred value of $b$ shifts away from zero and the uncertainty is further reduced, so that the deviation from the $\Lambda$CDM model becomes more pronounced in the full data combination. Thus, unlike the $f_1$CDM model, for which the inclusion of SNe Ia brings the result back into agreement with the $\Lambda$CDM model, the SNe Ia data strengthen the preference for a nonzero deformation parameter in the $f_2$CDM model.
This behavior shows that the response to the successive addition of CMB and SNe Ia information is strongly model dependent. In the $f_2$CDM case, the CMB distance priors reduce the deviation indicated by the BAO-only data, whereas the Pantheon+ SNe Ia measurements enhance it. However, a preference for a nonzero value of $b$ in the marginalized posterior does not necessarily imply that the $f_2$CDM model provides a statistically better description of the observations than the $\Lambda$CDM model. This point is examined below using the AIC and BIC.

The $f_3$CDM model shows a stronger dependence on the adopted data combination. The BAO-only analysis gives $n=1.86^{+0.07}_{-0.08}$ and $\Omega_m=0.22^{+0.03}_{-0.04}$, with a relatively broad degeneracy in the $n$--$\Omega_m$ plane. After including the CMB distance priors, the preferred parameter region changes substantially, with $n=1.60\pm0.01$ and $\Omega_m=0.30\pm0.01$. In particular, the preferred value of $\Omega_m$ shifts from about $0.22$ to $0.30$, accompanied by a clear shift in $n$ and a substantial reduction in the parameter uncertainties. The inclusion of the CMB distance information therefore changes not only the precision of the constraints, but also the location of the parameter region preferred by the BAO data.
The addition of the Pantheon+ SNe Ia sample produces a smaller further change. For the full combination, we obtain $n=1.62\pm0.01$, while $\Omega_m$ remains close to $0.31$. Therefore, unlike the cases of $f_1$CDM and $f_2$CDM models, where the main changes induced by the different data combinations can be discussed in terms of the departure from the $\Lambda$CDM model, the behavior of the $f_3$CDM model is characterized mainly by the sensitivity of its parameter space when different observations are combined. The results show that the BAO-only preferred region changes markedly after the CMB distance priors are added, whereas the subsequent inclusion of Pantheon+ produces a comparatively smaller shift.

The comparison among the three models shows that the inferred cosmological constraints are strongly dependent on the specific form of the $f(T)$ model, even when the same observational data combinations are used. This is particularly clear for the $f_1$CDM and $f_2$CDM cases, both of which can be directly compared with the $\Lambda$CDM model through the parameter $b$. For the $f_1$CDM model, the mild deviation obtained from BAO+CMB disappears after the Pantheon+ SNe Ia data are added, and the full combination becomes fully consistent with the $\Lambda$CDM model. In contrast, for the $f_2$CDM model, the BAO-only data already favor a positive value of $b$, the inclusion of the CMB distance priors reduces this deviation, and the addition of Pantheon+ strengthens it again. These different behaviors indicate that the departures from the $\Lambda$CDM model suggested by recent DESI BAO analyses are not reproduced in the same way for all cosmological models. In particular, the deviation found in dynamical dark energy analyses should not be interpreted as a generic consequence of the DESI BAO measurements alone. Its significance and even its direction can depend on the assumed form of the cosmological model and on how the different observations break the corresponding parameter degeneracies.
The $f_3$CDM model provides a complementary example because it cannot be compared with the $\Lambda$CDM model through a vanishing deformation parameter. In this case, the successive inclusion of the CMB distance priors and Pantheon+ SNe Ia data leads to clear changes in the allowed $n$--$\Omega_m$ parameter space. The BAO-only constraints leave a relatively broad degeneracy in the $n$--$\Omega_m$ plane and favor a substantially lower value of $\Omega_m$ than the joint constraints. After the CMB distance priors are added, both $n$ and $\Omega_m$ become much more tightly constrained and the preferred region moves considerably, whereas the subsequent addition of Pantheon+ produces a smaller further change. This behavior illustrates the different roles played by the three cosmological probes. The BAO data alone provide relatively weak constraints on the $n$--$\Omega_m$ parameter space, whereas the inclusion of the CMB distance priors significantly reduces the parameter degeneracy. The subsequent addition of the SNe Ia data produces only a relatively small further change in the constraints. The fact that the relative importance of these data sets differs among the three $f(T)$ models shows that current constraints on modified gravity are determined not only by the precision of the observations, but also by how each model changes the expansion history and how its parameters project onto the different cosmological measurements.

We next consider the transition redshift $z_T$, which provides a physically transparent characterization of the epoch at which the cosmic expansion changes from deceleration to acceleration \cite{Zhu:2003wv,2005ApJ...620....7Z,2010ApJ...711..439C,Kumar:2025mzo}. The expansion history can be characterized by the deceleration parameter
\begin{equation}
q(z)
\equiv
-\frac{\ddot{a}}{aH^2},
\end{equation}
where $a(t)$ is the scale factor and $H=\dot{a}/a$ is the Hubble parameter. In terms of the normalized Hubble parameter
\begin{equation}
E(z)\equiv\frac{H(z)}{H_0},
\end{equation}
the deceleration parameter can be written as
\begin{equation}
q(z)
=
\frac{1+z}{2E^2(z)}
\frac{dE^2(z)}{dz}
-1.
\end{equation}
The transition redshift is then defined by
\begin{equation}
q(z_T)=0,
\end{equation}
corresponding to the change from decelerated, $q>0$, to accelerated, $q<0$, cosmic expansion. For each posterior sample, we solve \(q(z_T)=0\), and quote the median and the 68\% credible interval of the resulting \(z_T\) distribution.

The derived transition redshift is relatively stable across the different models and data combinations. For the $f_1$CDM model, the constraints lie in the range $z_T\simeq0.62$--$0.67$, while the $f_2$CDM model gives $z_T\simeq0.65$--$0.70$. For $f_3$CDM, the BAO-only constraint, $z_T=0.70^{+0.14}_{-0.16}$, has a relatively large uncertainty, whereas the inclusion of the CMB distance priors and SNe Ia data gives $z_T\simeq0.67$--$0.68$ with substantially reduced uncertainties. As shown in Fig.~\ref{fig:zt}, the inferred values are also broadly consistent with the reference ranges from the Planck 2018 and Pantheon samples adopted in Ref.~\cite{Kumar:2025mzo}.
This behavior differs from that found for the model parameters. For the $f_1$CDM case, the preferred value of $b$ varies with the data combination and becomes consistent with the $\Lambda$CDM model after the SNe Ia data are included. For the $f_2$CDM model, the full data combination instead favors a nonzero value of $b$, while the allowed $n$--$\Omega_m$ region changes substantially after the CMB distance priors are added for the $f_3$CDM model. Despite these differences, the inferred values of $z_T$ vary only modestly. Therefore, the differences found in the parameter constraints do not lead to comparably large changes in the inferred epoch of the transition from decelerated to accelerated expansion.

Finally, we assess the relative statistical performance of the models by using the Akaike information criterion (AIC) \cite{AIC} and the Bayesian information criterion (BIC) \cite{BIC}, which account for both the goodness of fit and the number of free model parameters. They are defined as
\begin{equation}
\mathrm{AIC}
=
-2\ln\mathcal{L}_{\max}+2k,
\end{equation}
and
\begin{equation}
\mathrm{BIC}
=
-2\ln\mathcal{L}_{\max}+k\ln N,
\end{equation}
where $\mathcal{L}_{\max}$ is the maximum likelihood, $k$ denotes the number of free parameters, and $N$ is the number of data points included in the analysis. For the Gaussian likelihoods considered here, the likelihood contribution can be expressed in terms of the minimum chi-square when comparing models using the same data combination.

Using the $\Lambda$CDM model as the reference model, we define
\begin{equation}
\Delta\mathrm{AIC}
=
\mathrm{AIC}_{\mathrm{model}}
-
\mathrm{AIC}_{\Lambda\mathrm{CDM}},
\end{equation}
and
\begin{equation}
\Delta\mathrm{BIC}
=
\mathrm{BIC}_{\mathrm{model}}
-
\mathrm{BIC}_{\Lambda\mathrm{CDM}}.
\end{equation}
Accordingly, smaller values of the information criteria correspond to better statistical performance after accounting for model complexity. Values of $\Delta X\lesssim2$ generally indicate little statistical distinction between the models, whereas increasingly positive values indicate progressively stronger support for the reference $\Lambda$CDM model relative to the model being compared, where $X$ denotes either AIC or BIC.

The minimum chi-square values and information criteria are summarized in Table~\ref{tab:IC}. For the BAO-only data, the three $f(T)$ models yield relatively small differences in AIC and BIC with respect to the $\Lambda$CDM model. With the inclusion of the CMB distance priors, the $f_1$CDM model remains close to the $\Lambda$CDM model. For the $f_2$CDM model, the improvement in the minimum chi-square is negligible, giving $\Delta{\rm AIC}=1.75$ and $\Delta{\rm BIC}=2.53$. These values indicate little distinction according to AIC and a modest preference for the $\Lambda$CDM model according to BIC. In contrast, the $f_3$CDM model exhibits substantially larger information criterion differences.
For the full BAO+CMB+SN combination, the $f_2$CDM model yields a slightly lower $\chi^2_{\min}$ than the $\Lambda$CDM model. The improvement approximately compensates for the additional parameter in AIC, resulting in $\Delta\mathrm{AIC}=-0.10$, whereas BIC favors the $\Lambda$CDM model, with $\Delta\mathrm{BIC}=5.29$.
The $f_1$CDM model gives $\Delta\mathrm{AIC}=2.00$ and $\Delta\mathrm{BIC}=7.39$, while the $f_3$CDM model is more strongly disfavored, with $\Delta\mathrm{AIC}=19.17$ and $\Delta\mathrm{BIC}=24.56$.

Overall, the three $f(T)$ models show clearly different behaviors under the same observational combinations. The $f_1$CDM model becomes consistent with $\Lambda$CDM when the full BAO+CMB+SN data set is used, whereas the $f_2$CDM model retains a preference for a nonzero deformation parameter without a corresponding statistical preference over $\Lambda$CDM. The $f_3$CDM model shows a stronger dependence on the adopted data combination and a poorer statistical performance in the joint analysis. In contrast to these differences in the model parameters and statistical performance, the inferred transition redshift remains relatively stable among the three models. These results show that the constraints derived from DESI DR2 and the combined cosmological data depend significantly on the specific form of the $f(T)$ model.

\section{Conclusions}
\label{sec:conclusion}

Recent analyses based on DESI observations have reported indications of departures from the $\Lambda$CDM model within dynamical dark energy scenarios. In this work, we investigate whether similar behavior appears in three representative $f(T)$ cosmological models using the DESI DR2 BAO measurements, together with the CMB distance priors and Pantheon+ SNe Ia data. Our main conclusions can be summarized as follows.

First, the evidence for a deviation from the $\Lambda$CDM model is strongly dependent on the specific form of the $f(T)$ model and on the adopted data combination. For the $f_1$CDM model, the mild deviation found from BAO+CMB disappears after the Pantheon+ SNe Ia data are included, and the full combination is consistent with the $\Lambda$CDM model. In contrast, the $f_2$CDM model favors a positive and nonzero value of the deformation parameter $b$ in the full analysis. The $f_3$CDM model shows an even stronger dependence on the data combination, with a substantial change in the allowed $n$--$\Omega_m$ parameter space after the CMB distance priors are included. These results show that the departures from the $\Lambda$CDM model suggested by DESI analyses are not reproduced in a universal manner in $f(T)$ gravity.

Second, a deviation from the $\Lambda$CDM model at the parameter level does not necessarily imply a statistical preference for the corresponding modified gravity model. This is most clearly illustrated by the $f_2$CDM model. Although the full BAO+CMB+SN analysis favors a nonzero value of $b$ and gives a slightly lower minimum $\chi^2$ than the $\Lambda$CDM model, the improvement is not sufficient to establish a clear model selection preference once the additional degree of freedom is taken into account. The $f_1$CDM model also shows no statistical advantage over the $\Lambda$CDM model, while the $f_3$CDM model is more strongly disfavored by the combined data. The parameter constraints and information criterion results therefore provide complementary information when assessing possible departures from the standard cosmological model.

Third, the transition redshift remains comparatively stable despite the different behaviors of the model parameters. The three $f(T)$ models generally give $z_T$ values within a relatively narrow range and broadly consistent with the reference constraints from Planck 2018 and Pantheon sample. Thus, the differences in the modified gravity parameters do not lead to comparably large changes in the inferred epoch at which the Universe transitions from decelerated to accelerated expansion. 

The present analysis is based primarily on traditional observational datasets. A more complete test of $f(T)$ gravity will require observables that are directly sensitive to the growth of cosmic structure and gravitational perturbations. Future studies using independent late-time probes, such as gravitational-wave standard sirens, strong gravitational lensing, and fast radio bursts, together with observables sensitive to the growth of cosmic structure, will provide complementary tests of $f(T)$ gravity and help further distinguish between different model parameterizations. The increasing precision of upcoming large-scale-structure surveys will therefore provide an important opportunity to further test modified gravity as an alternative explanation of the late-time cosmic acceleration.

\acknowledgments
Zheng, J. would like to thank Prof. Jing-Zhao Qi of Northeastern University for valuable discussions and suggestions. Zheng, J. was supported by the National Natural Science Foundation of China under Grant Nos. 12403002 and 12433001.


\bibliographystyle{JHEP}
\bibliography{main.bib}


\end{document}